\documentclass[showpacs,twocolumn,preprintnumbers,amsmath,amssymb,pra]{revtex4-2}

\usepackage{graphicx}
\usepackage{nicefrac}
\usepackage{color}
\usepackage{subcaption}
\usepackage{ulem}

\usepackage{hyperref}
\usepackage{ulem}

\newcommand{\ket}[1]{\left| #1 \right\rangle}
\newcommand{\bra}[1]{\left\langle #1 \right|}

\newcommand{\braket}[2]{\left\langle #1 \middle| #2 \right\rangle}
\newcommand{\ketbra}[2]{\left|  #1 \middle\rangle \middle\langle #2 \right|}

\begin{document}

\title{ Measurement-induced cubic phase state generation}
% State classification for multimode nonlinear optical circuits

% repeat the \author .. \affiliation  etc. as needed
% \email, \thanks, \homepage, \altaffiliation all apply to the current
% author. Explanatory text should go in the []'s, actual e-mail
% address or url should go in the {}'s for \email and \homepage.
% Please use the appropriate macro foreach each type of information

% \affiliation command applies to all authors since the last
% \affiliation command. The \affiliation command should follow the
% other information
% \affiliation can be followed by \email, \homepage, \thanks as well.

\author{Harsh Kashyap$^{1}$}
\email{kashyapharsh16@gmail.com}
\author{Denis A. Kopylov$^{2,3}$}
\author{Polina R. Sharapova$^{2}$}
\email{polina.sharapova@upb.de}

\affiliation{
1. Department of Physical Sciences,
Indian Institute of Science Education and Research (IISER) Mohali, Sector 81, S.A.S. Nagar, Manauli PO 140306, India  \\ 
2. Department of Physics, Paderborn University, Warburger Str. 100, 33098 Paderborn, Germany\\
3. Institute for Photonic Quantum Systems (PhoQS), Paderborn University, Warburger Str. 100, 33098 Paderborn, Germany \\
}

\date{\today}

\begin{abstract}
The cubic phase state constitutes a nonlinear resource that is essential for universal quantum computing protocols. However, constructing such non-classical states faces many challenges. In this work, we present a protocol for generating a cubic phase state with high fidelity. The protocol is based on a  set of Gaussian operations assisted by a detection operation. To find the proper set of parameters that results in both high fidelity and high detection probability, we provide a numerical multiparameter optimization. We investigate a broad range of target states and study how parameter imperfections influence fidelity.  
\end{abstract}

% insert suggested keywords - APS authors don't need to do this
%\keywords{}

%\maketitle must follow title, authors, abstract, and keywords
\maketitle

%\tableofcontents

\section{Introduction}

Quantum optical circuits, where photons serve as carriers of quantum information and gates implement optical transformations, represent a promising platform for continuous-variable quantum computing \cite{Fukui}. However, to unlock the universality of quantum computing in continuous variables, in addition to the set of Gaussian staes and operations \cite{RevModPhys.84.621},  non-Gaussian resources are strongly required \cite{PhysRevLett.82.1784,  PhysRevResearch.6.023332}. 

%conventional quantum optical tools (linear optical elements, displacement, squeezing, etc.) are Gaussian \cite{RevModPhys.84.621} and therefore insufficient for universal quantum computing \cite{PhysRevLett.82.1784}. To unlock universality, in addition to the set of Gaussian operations, non-Gaussian resources are strongly required \cite{PRXQuantum.2.030204, PhysRevResearch.6.023332}.

Since the first ideas for constructing non-Gaussian states  ~\cite{Gottesman_2001, Ghose}, great progress has been made towards their theoretical description, classification,  characterization and experimental realization ~\cite{PRXQuantum.2.030204, PhysRevA.97.052317, PhysRevA.78.060303}.
Today's family of non-Gaussian states includes a broad set of GKP (Gottesman–Kitaev–Preskill) -generation proposals ~\cite{Brady} ranging from propagating-light demonstrations~\cite{doi:10.1126/science.adk7560} and linear optical circuits assisted by a set of photon-number-resolving (PNR) detectors~\cite{Tzitrin_2020}  to cavity-assisted preparation~\cite{PhysRevLett.128.170503} and free-electron-driven nonlinear interactions~\cite{PhysRevX.13.031001};  the generation of Schrödinger cat states by photon-subtraction ~\cite{PhysRevLett.97.083604, PhysRevLett.101.233605, Sychev} and photon-addition ~\cite{PhysRevA.110.023703} operations; the generation of squeezed cat states by leveraging adaptive measurement-based photonic circuits~\cite{Anteneh2024}; the use of Kerr nonlinearities ~\cite{Tyc_2008, Yao, Brauer}.

Among non-Gaussian resources, the cubic phase state ~\cite{Gottesman_2001} plays a special role, since it is necessary for the implementation of a cubic phase gate ~\cite{PhysRevLett.124.240503, PhysRevA.93.022301, Gu_2009, PhysRevApplied.15.024024}, the lowest-order non-Gaussian gate, which is one of the key nonlinear operations for universal continuous-variable quantum computing.
%it supports error-correction encoding and entanglement-distillation protocols~\cite{Gottesman_2001, Eisert_2003}. 
This state is characterized by nonlinear squeezing  ~\cite{Sakaguchi_2023, Kala_2025} and is attractive for error-correction protocols. However, in optics, the deterministic generation of high-fidelity cubic phase states is challenging because third-order optical nonlinearity is intrinsically weak. This has motivated probabilistic, measurement-induced strategies that approximate cubic states using experimentally accessible components and techniques, such as  squeezed state ancillas and photon subtraction
~\cite{PhysRevA.95.052352}, 
squeezed states accompanying by a proper sequence of photon subtractions and displacements
~\cite{PhysRevA.84.053802,  PhysRevA.88.053816}, squeezed states assisted by displacement and PNR detection ~\cite{PhysRevA.91.032321}, displaced
squeezed vacuum states, interferometers and PNR detection accompanied by machine-learning techniques for fidelity optimization
\cite{ PhysRevA.100.012326}.
Other approaches include protocols based on the Gaussian conversion of the trisqueezed states \cite{ Ferrini_2021}, the use of Kerr nonlinerarity assisted by proper squeezing and displacement ~\cite{PhysRevLett.124.240503}, deep reinforcement learning to control a quantum optical circuit for a near determenistic generation of cubic phase states with high success ~\cite{Anteneh2025}.

Recently, the generation of cubic phase states in the microwave domain has been demonstrated, opening new possibilities for the preparation of non-Gaussian resources.  For example, the cubic phase state preparation via three-wave mixing microwave
architecture based on the superconducting nonlinear asymmetric inductive element (SNAIL) resonator was theoretically predicted in ~\cite{PhysRevLett.125.160501} and experimentally realized in ~\cite{Eriksson2024NatComm}. The generation of cubic phase states based on the  sequence of interleaved selective number-dependent arbitrary phase (SNAP) gates and displacements was discussed in \cite{Kudra_2022}.

In this work, we introduce a new measurement-induced optical protocol for generating a cubic phase state. Our protocol  incorporates experimentally available Gaussian operations together with a single non-Gaussian projection onto the Fock state $\ket{2}$ and achieves fidelity of $0.97$ together with  the success probability of $0.27$ for a moderate squeezing ($\leq 6$~dB) after a proper optimization of the elements using the gradient-based optimization technique.  Our protocol demonstrates robustness to parameter drift of order $\sim 2\%$, making it practical for near-term tabletop implementations, and expands the toolbox for continuous-variable quantum computation.
Unlike previous approaches, our protocol is based on the Fock state $\ket{2}$ and the coherent state as initial states and utilizes only a single projection onto the Fock state $\ket{2}$, providing an experimentally accessible alternative to the existing protocols.

% We cast state preparation as an optimization problem and use gradient-based numerical methods to minimize infidelity between the heralded and target cubic states. With moderate squeezing ($\leq 6$~dB), we obtain fidelities above $0.97$ at success probabilities exceeding $0.5$, and we show robustness to $\sim 2\%$ parameter drifts. This renders the protocol practical for near-term tabletop implementations and expands the toolbox for CV quantum computation.  In contrast, our protocol uses only a single-shot projection onto the Fock state $\ket{2}$ together with standard Gaussian optics, yielding a compact, and experimentally easy alternative.}

\section{Theoretical Model}

%\subsection{Optical scheme under study}

\begin{figure}[ht]
\centering
\includegraphics[width=0.8\linewidth]{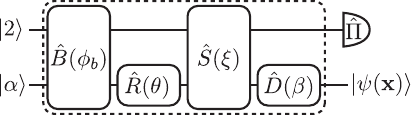}
\caption{
Optical circuit for the measurement-induced cubic-phase-state generation. $\hat{B}$: a beam splitter with an angle $\phi_{b}$; $\hat{R}$: a phase shifter with an angle $\theta$; $\hat{S}$: two-mode squeezing with  a squeezing parameter $\xi$;  $\hat{D}$: displacement with an amplitude $\beta$. The input state is a two-photon Fock state $\ket{2}$  and a coherent state $\ket{\alpha}$. A projection $\hat{\Pi}=\ketbra{2}{2}$ in the first output arm gives the cubic phase state in the second arm.}
\label{fig:scheme} 
\end{figure}

The studied optical scheme is depicted in Fig.~\ref{fig:scheme} and is based on the set of Gaussian operations (marked by the dashed rectangle) and subsequent post-selection via projection measurement.

At the input of the scheme, we use a two-mode state $\ket{\psi_0} = \ket{2}_1\otimes\ket{\alpha}_2$; i.e., the Fock state $\ket{2}$ in the first channel and the coherent seed $\ket{\alpha}$ with the real amplitude $\alpha$ in the second channel. 
The protocol we used consists of a beam splitter $\hat{B}(\phi_b)  = e^{i  \phi_b (\hat{a}_1^\dagger \hat{a}_2 + \hat{a}_1 \hat{a}_2^\dagger)} $ with an angle $\phi_b$, a phase shifter $\hat{R}_n(\theta ) = e^{i \theta \hat{a}_n^\dagger \hat{a}_n }$ with an angle $\theta$, a two-mode squeezer $\hat{S}(\xi) = e^{ \xi^* \hat{a}_1\hat{a}_2  - \xi \hat{a}_1^\dagger \hat{a}_2^\dagger  }$ with a complex squeezing parameter $\xi = |\xi| e^{i\phi_{\xi}}$ and a displacement operator $\hat{D}_n(\beta) =  e^{ \beta \hat{a}_n^\dagger - \beta^* \hat{a}_n }$ with a complex amplitude $\beta = |\beta| e^{i\phi_{\beta}}$, where $\hat{a}$ and $\hat{a}^\dagger$ are the annihilation and creation operators.
The lower indices $n = 1,2$ denote the channel number.
Note that a coherent state can be written with the use of a displacement operator $\ket{\alpha} = \hat{D}_2(\alpha)\ket{0}$, while an operator $\hat{D}(\alpha)$ can be represented as a part of  the considered Gaussian operations.  
Therefore, the  set of Gaussian operations performs a unitary transformation
$\hat{U}(\mathbf{x}) \equiv \hat{D}_2(\beta) \hat{S}(\xi) \hat{R}_2(\theta) \hat{B}(\phi_b) \hat{D}_2(\alpha)$ under the input state $\ket{\psi_0}\equiv\ket{\psi_0(0)}$, where $ \mathbf{x} \equiv (\alpha,\phi_b,\theta,|\xi|,\phi_{\xi},|\beta|,\phi_{\beta})$ is the vector of real parameters.
 % $\hat{U}(\mathbf{x}) = \hat{U}_n(x_n)  \cdot...\cdot  \hat{U}_1(x_1) $ under the input state, where $\hat{U}_i(x_i)$ is an $i$-th element of the interferometer, and  $ \mathbf{x} = [\alpha,\phi_b,\theta,|\xi|,\phi_{\xi},|\beta|,\phi_{\beta}]$ is the vector of real parameters.
At the output of the scheme we make a post-selection via a projection measurement $\hat{\Pi}$ in the first channel, and the output state reads
\begin{equation}
    \ket{\psi(\mathbf{x})} = \dfrac{\hat{\Pi} ~ \hat{U}(\mathbf{x}) \ket{\psi_{0}}}{ \mathcal{N}(\mathbf{x}) },
    \label{eq:output_state}
\end{equation}
where
$\mathcal{N}(\mathbf{x}) = \sqrt{ \bra{\psi_{0}} \hat{U}(\mathbf{x}) ~ \hat{\Pi} ~ \hat{U}(\mathbf{x}) \ket{\psi_{0}  }  }$ is the normalization coefficient.
In this paper, we limit ourself to the case of $ \hat{\Pi} = \ketbra{2}{2}$.
A choice of such projection measurement, as well as the initial states, is based on the photon-number representation of the cubic phase state for small values of cubicity.

As a target state we use the cubic phase state \cite{Gottesman_2001}
\begin{equation}
    \ket{\psi_{T}} = e^{ir\hat{q}^3}\hat{S}(\xi_{T})\ket{0},
    \label{eq:cubic_phase}
\end{equation}
where $r$ is the strength of the cubic interaction known as cubicity, $\hat{q} = \frac{\hat{a} + \hat{a}^\dagger}{2}$ is the position quadrature and $\xi_{T}$ is the squeezing parameter. 
The squeezing parameter can be defined as $\xi_{T} = -\ln[10^{\xi_{dB}/20}]$, where $\xi_{dB}$ is the squeezing degree in the dB-scale.

The task of generating the target state can be formulated as the optimization task of finding the optimal set of parameters $\mathbf{x}_0$ that minimizes a loss function $\mathcal{L}(\mathbf{\mathbf{x}_0}$). 
As a loss function we use the infidelity
\begin{equation}
    \mathcal{L}(\mathbf{x}) = 1 - \mathcal{F}(\mathbf{x}),
    \label{eq:infidelity}
\end{equation}
where the fidelity $\mathcal{F}(\mathbf{x}) = |\braket{\psi(\mathbf{x}) }{  \psi_{T}}|^2 $. 
To find the optimal set of parameters for the studied system, we use the gradient-based optimization technique. 
In Appendices~\ref{app_gradients_optimization} and \ref{app_operators}, we show how the gradients for the studied operators can be calculated. In Appendix~\ref{app_optimization_realization}, we describe in detail the optimization protocol used.

\section{Results and Discussion}
\subsection{Single state Optimizations}
 
\begin{figure}[ht]
    \includegraphics[width=1.\linewidth]{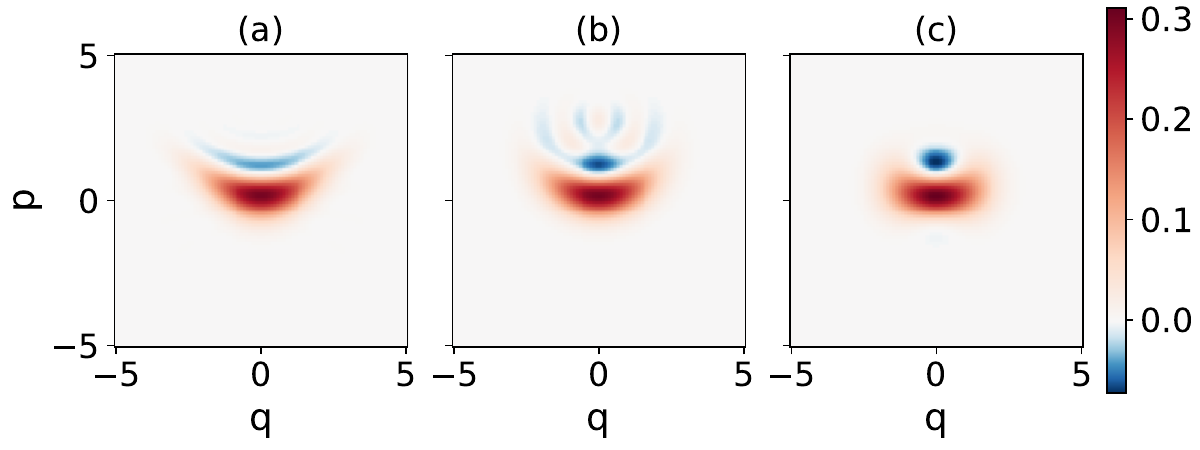}
    \caption{Wigner functions for (a) the ideal cubic phase state $\ket{\psi_{T}}$ with $r = 0.15$ and $\xi_{dB}=5$ dB and (b,c) the states $\ket{\psi_{out}}$ generated in the considered protocol. In (b) all protocol parameters are optimized (see Target 1 in TABLE~\ref{table_1}), while in (c) the beam splitter angle is fixed as $\phi_b = \frac{\pi}{4}$, all other protocol parameters are optimized (see Target 1 in TABLE~\ref{table_2}). The Wigner functions were calculated with the use of QuTiP software~\cite{QuTiP_2024}.}
    \label{fig:fig_2_wigner}
\end{figure}

%An initial study was performed on some specific \pscomment{specific?} target cubic phase states. 
In Fig.~\ref{fig:fig_2_wigner}, we show examples of how our protocol operates.
 Fig.~\ref{fig:fig_2_wigner}a presents the Wigner function of the ideal cubic phase state with cubicity $r = 0.15$ and squeezing strength $\xi_{dB}=5$ dB. 
In Fig.~\ref{fig:fig_2_wigner}b, we show the Wigner function of the output state generated in our protocol, when optimizing the  scheme parameters in order to get the same cubic phase state as in Fig.~\ref{fig:fig_2_wigner}a, the optimized parameters for this target state along with two other target states are given in TABLE~\ref{table_1}.  
One can notice that the protocol results in a very high fidelity between the target and the generated states $F=0.9933$, however, to reach such fidelity, the beam splitter angle $\phi_b$ should be very close to $\frac{\pi}{2}$ meaning the almost full intensity reflection. In addition, the optimal $\alpha$ value should be very small that significantly affects the detection probability, making it of the order of $10^{-8}$, which, in turn, complicates the experimental implementation of the scheme. Thus, TABLE~\ref{table_1} represents a theoretical upper bound on the fidelity that can be achieved using highly unbalanced beam splitters and very small coherent amplitudes.

%This also affects the optimal $\alpha$ value, making it very small. 
%In this case, the detection probability is of the order of $10^{-7}$, which makes the scheme difficult to realize experimentally.

%The same behaviour can be seen for  other target states in TABLE~\ref{table_1}.

\begin{table}[htbp]
\begin{ruledtabular}
\begin{tabular}{cccc}
 & Target 1 & Target 2 & Target 3 \\
\hline
\hline
$r$ & 0.15 & 0.20 & 0.15 \\
$\xi_{dB}$ [dB] & 5 & 5 & 6 \\
\hline
$\alpha$ & 0.0100 & 0.0100 & 0.0100 \\
$\phi_b$ & $0.49856\pi$ & $0.50119\pi$ & $0.49880\pi$ \\
$\theta$ & $0.50000\pi$ & $1.50000\pi$ & $0.50000\pi$ \\
$|\epsilon|$ & 0.01073 & 0.00959 & 0.01160 \\
$\phi_{\epsilon}$ & $1.50000\pi$ & $0.50000\pi$ & $1.50000\pi$ \\
$|\beta|$ & 1.5725 & 1.6250 & 1.6200 \\
$\phi_{\beta}$ & $1.50000\pi$ & $1.50000\pi$ & $1.50000\pi$ \\
\hline
\hline
Fidelity & 0.9933 & 0.9912 & 0.9861 \\
\hline
\hline
Detection \\ Probability & $5.0223\times 10^{-8}$  & $3.7457\times 10^{-8}$  &  $5.0368\times 10^{-8} $ \\
\end{tabular}
\end{ruledtabular}
\caption{ Optimized  parameters for different target states. }
\label{table_1}
\end{table}

%From the TABLE~\ref{table_1}, we can see that, for $r = 0.15$ and $sq = 5$~dB, $\phi_b = 1.5706 \approx \frac{\pi}{2}$, which means almost zero transmission intensity, $T = 0$ and full reflection intensity of the beam splitter, which also affects the $\alpha$ value, which is very small in this case. ANd same behaviour can be seen in the other target states as well, which are given in the TABLE~\ref{table_1} 

To avoid low detection probability and examine experimentally relevant regime, we consider the balanced beam splitter by setting its angle at $\phi_b = \frac{\pi}{4}$ and optimizing all other protocol parameters. This allows us to increase the detection probability, as well as the amplitude of the initial coherent seed, while maintaining the fidelity high enough $F=0.9735$, see TABLE~\ref{table_2}. The plot of the Wigner function of such optimized state is presented in Fig.~\ref{fig:fig_2_wigner}c.

%set the beam splitter at transmission intensity of $50\$, $T = 0.5$ , which results in improved $\alpha$ values as shown in Fig.~\ref{fig:prob_and_alpha} and TABLE~\ref{table_2} for the same target states. Wigner function plot for the optimized state of Target 1 is given in Fig.~\ref{fig:combined r = 0.15, sq = 5}c.

\begin{table}[htbp ]
% \begin{ruledtabular}
% \begin{tabular}{cccc}
%  &Target 1&Target 2&Target 3 \\ 
% \hline
% \hline
% r & 0.15 & 0.2 & 0.15 \\
% sq[dB] & 5 & 5 & 6 \\
% \hline
% $\alpha$ & 0.2202 & 0.1192 & 0.2270 \\
% $\theta$ & 0.49999$\pi$ & 0.49999$\pi$ & 0.49999$\pi$ \\
% $|\xi|$ & 0.1293 & 0.1459    & 0.1566 \\
% $\phi_{\xi}$ & 0.99999$\pi$ & 0.99999$\pi$ & 0.99999$\pi$ \\ 
% $|\beta|$ & 0.1814 & 0.2390 & 0.1805 \\
% $\phi_{\beta}$ & 1.50000$\pi$ & 1.50000$\pi$ & 1.50000$\pi$ \\
% \hline
% \hline
% Fidelity & 0.9735 & 0.9683 & 0.9532\\ 
% \hline
% \hline
% Detection \\
% Probability& 0.5170 & 0.5130 & 0.5246
% \end{tabular}
% \end{ruledtabular}
\begin{ruledtabular}
\begin{tabular}{cccc}
 &Target 1&Target 2&Target 3 \\ 
\hline
\hline
r & 0.15 & 0.2 & 0.15 \\
$\xi_{dB}$[dB] & 5 & 5 & 6 \\
\hline
$\alpha$ & 0.2202 & 0.1192 & 0.2270 \\
$\theta$ & 0.5$\pi$ & 0.5$\pi$ & 0.5$\pi$ \\
$|\xi|$ & 0.1293 & 0.1459    & 0.1566 \\
$\phi_{\xi}$ & $\pi$ & $\pi$ & $\pi$ \\ 
$|\beta|$ & 0.1814 & 0.2390 & 0.1805 \\
$\phi_{\beta}$ & 1.5$\pi$ & 1.5$\pi$ & 1.5$\pi$ \\
\hline
\hline
Fidelity & 0.9735 & 0.9683 & 0.9532\\ 
\hline
\hline
Detection \\Probability 
 & 0.2673 & 0.2632 & 0.2752
\end{tabular}
\end{ruledtabular}

% Temporarily override caption settings for this table
% \captionsetup{justification=centering,singlelinecheck=true}
\caption{Optimized  parameters for different target states. The beam splitter angle is fixed at $\phi_b = \frac{\pi}{4}$. }
\label{table_2}
\end{table}

% \added{\noindent\textbf{Remark (Tables I vs.~II).} Table~\ref{table_1} exhibits the \emph{performance bound}: allowing all parameters to vary yields $F\gtrsim 0.99$ but extremely small success probabilities ($\sim 10^{-7}$) associated with highly unbalanced beam splitting ($\phi_b\approx \pi/2$). Table~\ref{table_2} fixes a balanced splitter ($T=0.5$), giving slightly lower fidelity ($F\approx 0.97$) but orders-of-magnitude higher success probability ($\sim 0.5$). Thus, Table~\ref{table_1} is a theoretical upper limit, while Table~\ref{table_2} reflects experimentally realistic operation.}

\subsection{Multiple state Optimizations}
 
In this section, we extend 
our previous study
%the analytics \pscomment{Did we perform analytics ?} 
to multiple target states by forming a grid, where each point on the grid corresponds to a unique target state.
To find a set of parameters that minimizes the loss function for each target state, we use the numerical continuation method, in which the optimized set of parameters for the fixed starting state is used to optimize parameters for the neighboring state close to the starting one.
%on the condition that the difference $\Delta$ between the state is small enough
This method is outlined in Appendix \ref{app_optimization_realization} as the second strategy.
The performed continuous optimization requires less time compared to brute-force optimization with generation of a large number of initial random states (the first strategy in Appendix \ref{app_optimization_realization}) and results in the same fidelity values, see Fig.~\ref{fig:Violin plot} in Appendix \ref{app_optimization_realization}.
 In the following, as the starting point for optimization, we choose a target state with parameters  $r = 0.15$ and $\xi_{dB}=5$~dB. 
 
 %The efficiency of our method is given in the Fig.~\ref{fig:Violin plot} in the Appendix, from which it is visible that our method is efficient \pscomment{How we can  see that our method is efficient ?} \Harsh{I have written it in the caption of the Fig. 6 should I add something in the main text}, and it give us the best fidelity for multiple target states.

 %\pscomment{What do we mean by the seed?}\Harsh{seed means that this point r=  0.15, sq = 5 dB is the state for which we do the optimization independently by using 100 initial conditions, and after we get the optimized solution for this target state, then we use the optimized values of the parameters to find the optimized value for the next state, which is $r+\Delta r$, $sq+\Delta sq$}, reason being we need to have seed point which has good fidelity but not at the edge of the grid.

\begin{figure}[ht]
    \centering
    \includegraphics[width=\linewidth]{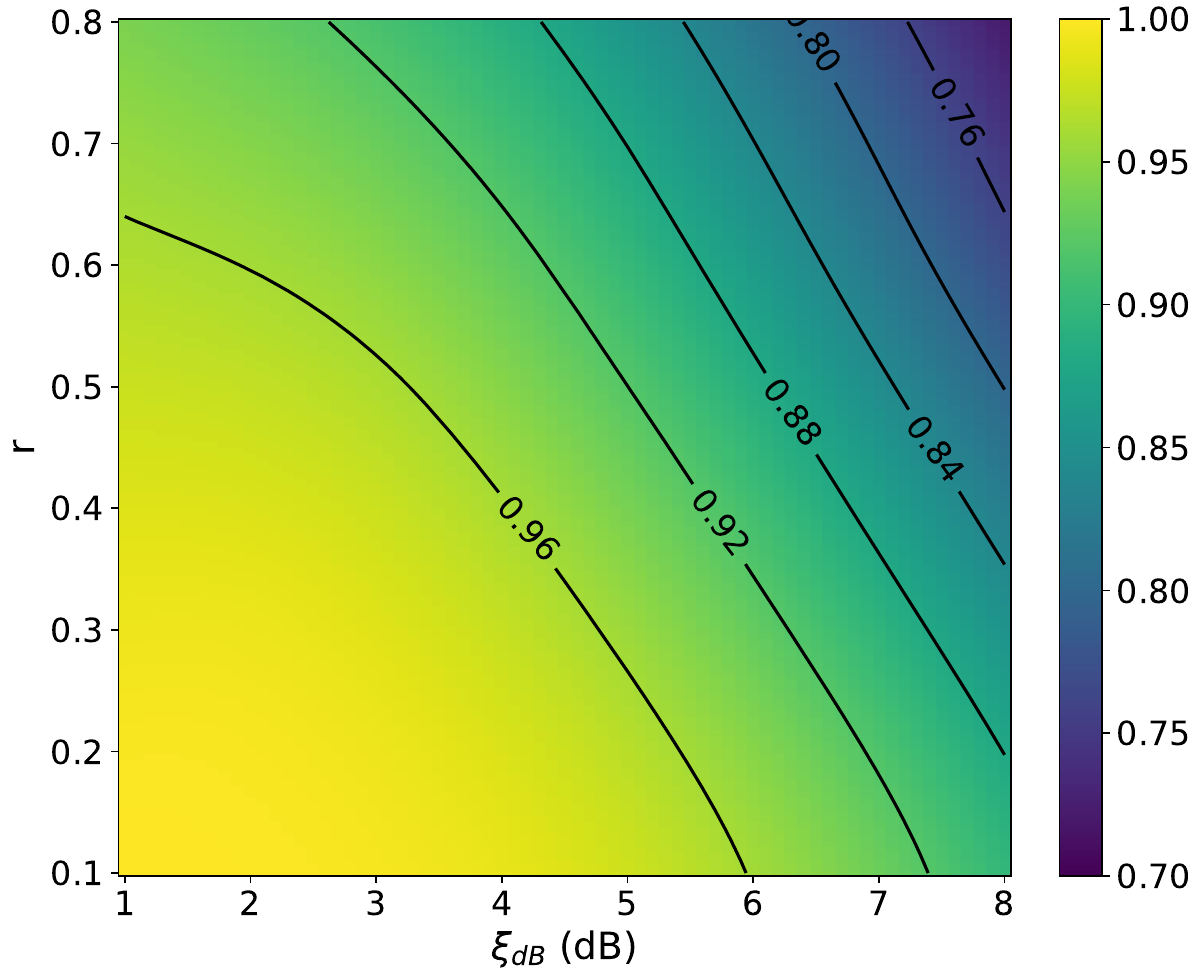}
    \caption{Optimized fidelity  for different combinations of $r$ and $\xi_{dB}$ values (target states), $T = 0.5$. }
    \label{fig:fig_3}
\end{figure}

The fidelity to generate cubic phase states with various values of cubicity and squeezing in our protocol is presented in Fig.~\ref{fig:fig_3}. Here, we used the multiple target states optimization by implementing the numerical continuation technique mentioned above. It can be seen that the presented protocol works better for smaller values of squeezing and cubicity, which is due to the fixed projection measurement onto the two-photon state. Indeed, as the squeezing and cubicity increase, the photon statistics of the cubic phase state becomes more complex and requires more advanced protocols involving projections to higher Fock states.

In Fig.~\ref{fig:fig_3}, the beamsplitter angle is fixed as $\phi_b=\pi/4$ (transmission coefficient $T=0.5$) to achieve a reasonable detection probability. However, the beamsplitter angle can be considered as a flexible parameter:  As shown in Fig.\ref{fig:comparison}a in Appendix~\ref{app_comparison} the transmission coefficient $T=0.8$ leads to similar results as in  Fig.~\ref{fig:fig_3}.

%angle  We also did the same thing by making the beam splitter at transmission intensity of 80\%. And the result is given in Appendix~\ref{app_comparison}

\begin{figure}[ht]
    \centering    \includegraphics[width=\linewidth]{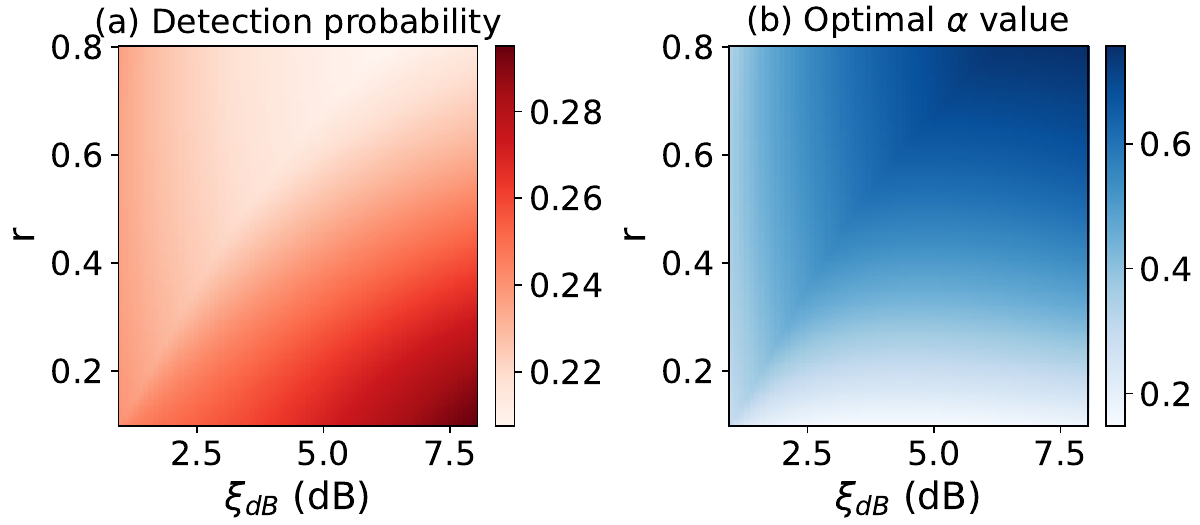}
    \caption{For  different target states, (a) detection probability in the upper channel and  (b) optimized $\alpha$ values. $T = 0.5$ is fixed. }
    \label{fig:prob_and_alpha}
\end{figure}

The detection probability $\mathcal{N}(\mathbf{x}) = \sqrt{ \bra{\psi_{0}} \hat{U}(\mathbf{x}) ~ \hat{\Pi} ~ \hat{U}(\mathbf{x}) \ket{\psi_{0}  }  }$ for the projection onto the two-photon state $ \hat{\Pi} = \ketbra{2}{2}$ that corresponds to Fig.~\ref{fig:fig_3} is shown in Fig. \ref{fig:prob_and_alpha}(a). The detection probability  deviates slightly over the entire range of the considered target states, but has a maximum for large squeezing values, demonstrating a trade-off between high fidelity and high detection probability. Fig. \ref{fig:prob_and_alpha}(b) presents the optimal values of the initial coherent state amplitude $\alpha$ (which is supposed to be real) corresponding to Fig.~\ref{fig:fig_3}. Here, as expected, to realize a cubic state with larger values of squeezing and cubicity, a larger amplitude of the initial coherent state is required, since the second initial state is fixed as a two-photon state.

\begin{figure}[htbp]
    \includegraphics[width=1.\linewidth]{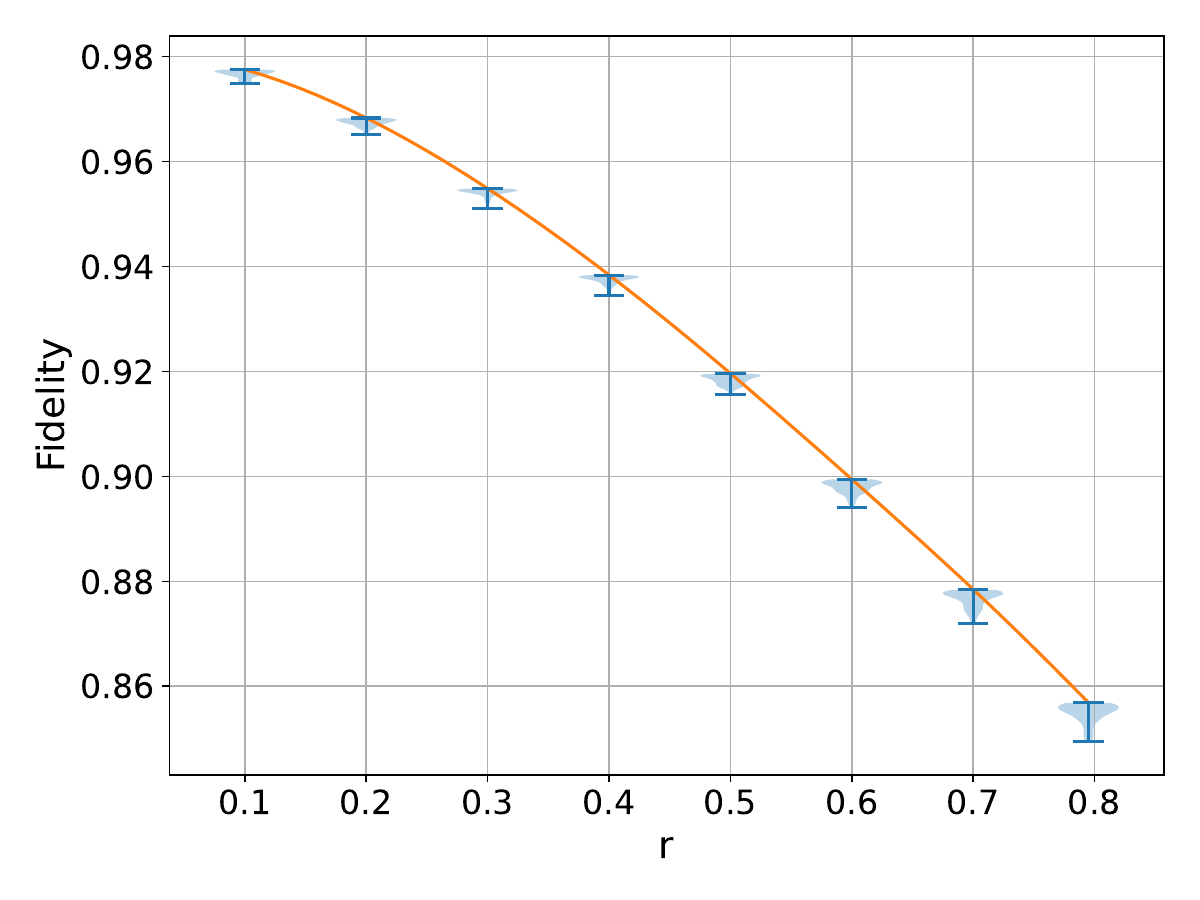}
    \caption{Stability of the method used. The orange line indicates the fidelity values  obtained using the continuation method when $T = 0.5$. 
    The violin plots at different 
    $r$ values illustrate the distribution of fidelity when an error of up to $2\%$ is introduced in the parameters.}
    \label{fig:error_plot_2_percent} 
\end{figure}

 To demonstrate the stability of the presented protocol, %parameters obtained from the optimization protocol, 
 we calculated the fidelity for randomly introducing  up to $2\%$ error in each optimized parameter found via the numerical continuation technique in Fig.~\ref{fig:fig_3}. Fig.~\ref{fig:error_plot_2_percent} presents such fidelity over the cubicity range for the fixed squeezing parameter of $\xi_{dB}=5$ dB. 
  For each $r$ value, the random error generation process  is performed 50 times, the obtained fidelity values are shown as a shaded violin-shaped area.  The spread of each violin area indicates the deviation in fidelity under the error introduced. It can be seen that states with higher cubicity values are more sensitive to the instability of optical elements and require more careful experimental realization.  
 % \pscomment{Do we know which element is more crucial for errors ? }

\section{Conclusion}

We presented a protocol for generating cubic phase states based on the set of unitary operations and projection measurement. Using the multiparameter optimization technique, we found the set of optimal parameters resulting in high fidelity and high detection probability for a set of target states. 
 To highlight the influence of errors  on the system performance, we estimated the stability of the presented protocol by introducing an error into each optical element, making the proposal suitable for the experimental realization.
It should be noted that all presented elements of the protocol are well-studied and experimentally feasible. For example, the experimental generation of the two-photon Fock state was realized in \cite{Cooper, Bimbard}, while the interference of the Fock state and a coherent state on a  beam splitter has been widely studied theoretically \cite{Hu2016, Xu2014, Xu2012} and  experimentally realized for the single-photon state and a variable beamsplitter \cite{PhysRevA.86.043820}. A dynamical in-line squeezing gate was experimentally implemented in \cite{PhysRevA.90.060302}, an in-line state squeezing using the seeded parametric down-conversion was realized in  \cite{Wang2022}, a scheme of the measurement-induced two-mode squeezing was theoretically proposed in \cite{Riabinin}.  Projective $\ketbra{2}{2}$ measurement can be performed using photon-number-resolving superconducting nanowire detectors that already demonstrate the system detection efficiencies above 90\%  \cite{Stasi_2023}.
%\added{Recent developments in photon-number-resolving superconducting nanowire detectors have significantly improved the feasibility of implementing projective measurements onto low-photon-number Fock states such as $\ketbra{2}{2}$. In particular, parallel SNSPD architectures now achieve system detection efficiencies above 90\%, and, crucially, non-Gaussian discrimination of 1-, 2-, 3- and 4-photon events with quantified fidelity, including explicit discrimination of two-photon events with fidelity $ P_{22} = 0.487$ has been demonstrated using parallel superconducting nanowire detectors \cite{Stasi_2023} }

\begin{acknowledgments}
    This work is supported by the  `Photonic Quantum Computing'  (PhoQC) project, funded by the Ministry for Culture and Science of the State of North-Rhine Westphalia.
     We also acknowledge financial support of the Deutsche Forschungsgemeinschaft (DFG) via the TRR 142/3
(Project No. 231447078, Subproject No. C10).
\end{acknowledgments}

\newpage
~
\newpage

\appendix

\section{Gradient-based optimization}
\label{app_gradients_optimization}

For the implementation of the gradient-based optimization, the gradients $\nabla_\mathbf{x} \mathcal{L}(\mathbf{x})$ of loss function Eq.~\ref{eq:infidelity} should be computed for all intermediate steps.
The $i$-th component of the gradient reads 
 \begin{equation}
    \dfrac{\partial \mathcal{L}(\mathbf{x}) }{\partial x_i}  = - 2 \mathrm{Re} \Big[ \braket{ \psi(\mathbf{x})  }{  \psi_T} \braket{\psi_T }{ \partial_{x_i}  \psi(\mathbf{x}) } \Big],
    \label{eq:loss_function}
\end{equation}
where the symbol $\ket{\partial_\lambda \psi }$ denotes the ket-vector $\ket{\partial_\lambda \psi }\equiv \frac{\partial \ket{\psi}  }{\partial \lambda}$.
% In the similar way, the Hessian matrix can be calculated as $h_{ij} = \partial^2 \mathcal{L}(\mathbf{x})/\partial x_i \partial x_j$.
In turn, the derivatives of the output state are
\begin{equation}
      \ket{\partial_{x_i} \psi(\mathbf{x})} = \dfrac{\partial }{\partial x_i} \bigg[  \dfrac{\hat{\Pi} ~   \hat{U}(\mathbf{x}) \ket{\psi_{0}}}{\mathcal{N}(\mathbf{x})} \bigg].
      \label{eq_state_derivative}
\end{equation}

Let us consider the intermediate state $\ket{\Psi}$ after all unitary operations, but before the detection, namely $\ket{\Psi} = \hat{U}(\mathbf{x}) \ket{\psi_{0}}$. 
Then, the derivative for the output state reads 
\begin{equation}
    \ket{\partial_{x_i} \psi(\mathbf{x})} = \dfrac{ 1 }{\mathcal{N}(\mathbf{x})} \hat{\Pi} \ket{ \partial_{x_i} \Psi }  - \frac{A_i }{2(\mathcal{N}(\mathbf{x}))^3} \hat{\Pi} \ket{\Psi } ,
    \label{eq:derivative_projection}
\end{equation}
where $\mathcal{N}(\mathbf{x}) = \sqrt{ \bra{\Psi} \hat{\Pi} \ket{\Psi } } $ and  $A_i = 2\mathrm{Re} \bra{\Psi }\hat{\Pi}\ket{\partial_{x_i} \Psi}$.

The major computational bottleneck for the gradients $\ket{ \partial_{x_i} \Psi }$ arises from the dependence of the operator $\hat{U}(\mathbf{x})$ on the vector $\mathbf{x}$.
The numerical differentiation via finite differences is inefficient: it needs additional computation of the operators $\hat{U}_{\Delta x_i}(\mathbf{x}+\Delta x \mathbf{e}_i)$ for all the parameters $\mathbf{x}$, where $\mathbf{e}_i = (0, ... , 1_i, ... 0,  )$ is a unit vector along $i$-th component.
However, for the studied optical scheme, it is possible to calculate the gradients $\partial_{x_i}\hat{U}(\mathbf{x})$ at the point $\mathbf{x}_0$ using only the operators $\hat{U}(\mathbf{x}_0)$.
The explicit expressions for the studied operators are given in Appendix~\ref{app_operators}.

\section{Gradients for Gaussian operators}
\label{app_operators}

The derivative of the phase-shift operator implemented in the $j$-th channel $\hat{R}_j(\theta) = e^{i \theta\hat{n}_j}$ reads
\begin{equation}
\dfrac{\partial\hat{R}_j(\theta)}{\partial\theta} = i \hat{n}_j\hat{R}_j(\theta), 
\end{equation}
where $\hat{n}_j = \hat{a}_j^\dagger \hat{a}_j $ is the photon-number operator. 

A similar expression can be obtained for the beamsplitter operator $\hat{B}(\phi_b ) = e^{i  \phi_b (\hat{a}_1^\dagger \hat{a}_2 + \hat{a}_1 \hat{a}_2^\dagger)}$:
\begin{equation}
    \dfrac{\partial\hat{B}(\phi_b )}{\partial\phi_b } = i (\hat{a}_1^\dagger \hat{a}_2 + \hat{a}_1 \hat{a}_2^\dagger)\hat{B}(\phi_b ), 
\end{equation}
where $\hat{a}_1$ and $\hat{a}_2$ are the annihilation operators in the first and second channels, respectively.

The derivative of the  two-mode squeezing operator $\hat{S}(\xi) = e^{ \xi^* \hat{a}_1\hat{a}_2  - \xi \hat{a}_1^\dagger \hat{a}_2^\dagger  }$ is a little more complicated because 
the squeezing parameter  is complex, namely $\xi = |\xi|e^{i\phi_{\xi}}$. This means that the derivative should be taken for both real amplitude $|\xi|$ and phase $\phi_{\xi}$.
To do this, we rewrite 
the squeezing operator  in a form where the phase and amplitude dependencies are separated  \cite{Rhodes_1990}:
\begin{equation}
    \hat{S}(\xi) = \big(\hat{R}^\dagger_1(\phi_{\xi})\otimes\hat{R}^\dagger_2(\phi_{\xi})\big) \hat{S}(|\xi|) \big(\hat{R}_1(\phi_{\xi})\otimes\hat{R}_2(\phi_{\xi})\big).
    \label{eq:separation}
\end{equation}
Then the derivative with respect to the amplitude $|\xi|$ reads 
\begin{multline}
    \dfrac{\partial\hat{S}(\xi)}{\partial |\xi|} =  \big(\hat{R}^\dagger_1(\phi_{\xi})\otimes\hat{R}^\dagger_2(\phi_{\xi})\big)  \\ \cdot ( \hat{a}_1\hat{a}_2  -  \hat{a}_1^\dagger \hat{a}_2^\dagger ) \hat{S}(|\xi|) \big(\hat{R}_1(\phi_{\xi})\otimes\hat{R}_2(\phi_{\xi})\big),
\end{multline}
while the gradient with respect to the phase $\phi_{\xi}$ is given by
\begin{equation}
    \dfrac{\partial\hat{S}(\xi)}{\partial \phi_{\xi}} = i[\hat{S}(\xi), \hat{n}_1\otimes \hat{n}_2 ],
\end{equation}
where the brackets $[. ~,.]$ denote the commutator.

Similar to \ref{eq:separation} expression can be written for the displacement operator $\hat{D}(\beta) =  e^{ \beta \hat{a}^\dagger - \beta^* \hat{a} }$ with a complex amplitude $\beta= |\beta|e^{i\phi_\beta}$:
\begin{equation}
    \hat{D}(\beta) = \hat{R}^\dagger(\phi_\beta)  \hat{D}(|\beta|) \hat{R}(\phi_\beta),
\end{equation}
which leads to the derivatives 
\begin{equation}
    \dfrac{\partial\hat{D}(\beta)}{\partial |\beta|} = \hat{R}^\dagger(\phi_\beta)  ( \hat{a}^\dagger - \hat{a} ) \hat{D}(|\beta|) \hat{R}(\phi_\beta),
\end{equation}
and
\begin{equation}
    \dfrac{\partial\hat{D}(\gamma)}{\partial \phi_\gamma} =  i [\hat{D}(\gamma), \hat{n}] .
\end{equation}

\section{ Optimization protocol and its realization  }
\label{app_optimization_realization}

In this paper, our computations are based on the truncated Fock state representation, therefore for the numerical optimization we have chosen the L-BFGS-B algorithm from the scipy library \cite{2020SciPy}.
To calculate the gradients of the loss function at each point $\mathbf{x}_n = (\alpha_n,\phi_{b_n},\theta_n,|\xi|_n,(\phi_{\xi})_n, |\beta|_n,(\phi_{\beta})_n)$, we implement the following algorithm:
\begin{enumerate}
    \item Compute the matrices 
        $\hat{D}(\beta_n)$, $\hat{S}(\xi_n)$, $\hat{R}(\theta_n)$, $\hat{B}(\phi_{b_n})$, $\hat{D}(\alpha_n)$ 
    and the $\partial_{x_i} \hat{U}(\mathbf{x}_n)$ using the corresponding equations from Appendix~\ref{app_operators}.
    % This can be done either by the direct computation of matrix exponent
    %  or via the iterativive method, presented in Ref.~[?].
    %  and $\partial_{x_i} \hat{U}_i(x_i) = \mathcal{M}_i [\hat{U}_i(x_i)] $. 
    % \item For $i$-th element with the use of Eqs.~\eqref{eq:output_state},~\eqref{eq:derivative_unitary} and~\eqref{eq:derivative_projection} compute the gradient of loss function Eq.~\eqref{eq:loss_function} by mutlipying corresponding matrices from $\{ \hat{U}_i(x_i) \}$ and $\{\partial_{x_i} \hat{U}(\mathbf{x}_n) \}$ .
    \item Find the next point $\mathbf{x}_{n+1}$ using the gradient-descend method.
\end{enumerate}

However, the gradient-based optimization provides the finding of a local minimum.
In order to find a global one, we perform two strategies. 
The first is to find the local minima for the fixed target state $\ket{\psi}_T$ with the randomly generated initial vectors $\mathbf{x}^i$.
In this case, the parameters corresponding to the lowest loss function are assumed to be the global minimum. %\pscomment{Do you mean to generate N initial random states ?} \Harsh{,  N = 100}

The second strategy assumes that for one fixed target state $\ket{\psi(\mathbf{p})}_T$ we know a set of parameters $\mathbf{x}_\mathbf{p}$ corresponding to the global minimum of the  loss function $\mathcal{L}(\mathbf{x}_\mathbf{p}) = 1 - |\braket{\psi(\mathbf{x}_\mathbf{p}) }{  \psi(\mathbf{p})}_T|^2$. This set can be, for example, found using the first strategy.
%is applied for the target state which depends on the parameters $\mathbf{p}$.
%Let us assume that we have a target state $\ket{\psi(\mathbf{p})}_T$ and we know the parameters $\mathbf{x}_\mathbf{p}$, where the loss function has a global minimum $\mathcal{L}(\mathbf{x}_\mathbf{p}) = 1 - |\braket{\psi(\mathbf{x}_\mathbf{p}) }{  \psi(\mathbf{p})}_T|^2$.
Then, for a target state $ \ket{\psi(\mathbf{p}+\mathbf{\Delta p})}_T $ with parameters shifted by a small step $\mathbf{\Delta p}$, the optimal parameters of the setup $\mathbf{x}_{\mathbf{p}+\mathbf{\Delta p}}$ corresponding to a global minimum of the loss function can be found using $\mathbf{x}_p$ as an initial parameter in the gradient descent algorithm.
% Therefore, the optimal parameters $\mathbf{x}_{\mathbf{p}+\mathbf{\Delta p}}$

%   \pscomment{And how do we find $\mathbf{x}_{\mathbf{p}+\mathbf{\Delta p}}$ ?}
% % This procedure is known as \DK{?? [?]}

\begin{figure}[htbp]
    \includegraphics[width=1.\linewidth]{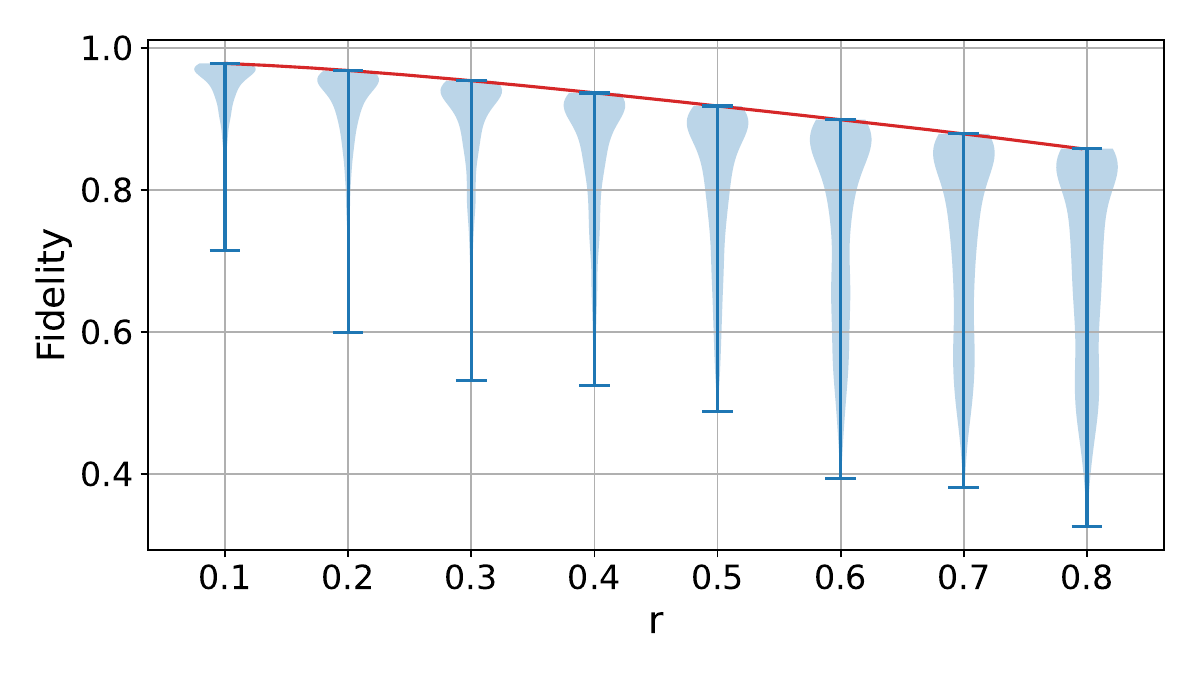}
    \caption{Efficiency of the method used. The red line indicates the fidelity values  obtained using the continuation method (second strategy) when $T = 0.8$. 
    The violin plots represent the distribution of the fidelity obtained for the independent optimization of the randomly generated initial states (first strategy) for the fixed r-value. The blue lines depicts the fidelity range achieved, the shaded blue are shows how frequently the fixed value of fidelity was obtained.}
    \label{fig:Violin plot} 
\end{figure}

In Fig.~\ref{fig:Violin plot}, we show a comparison of the fidelity values obtained using the two strategies, depending on the cubicity $r$. The squeezing parameter of the target state and the transmission coefficient of the beamsplitter are fixed as $\xi_{dB}=5$~dB and $T = 0.8$, respectively.
For the first strategy, the fidelities were calculated for $N=100$ random initial vectors. The distribution of the obtained local minima is depicted by the violin plots (blue-shaded area). It can be seen that the largest amount of fidelity values is localized near the red solid line. 
To realize the second strategy, we used the best (optimal) parameters of the scheme found for $r=0.15$ in the first strategy. 
Here, changing the parameter $r$ with the step of 0.005, we have obtained the optimal fidelity values for all other cubicities of the target state (red line in Fig.~\ref{fig:Violin plot}).  
One can notice that the maximal fidelity obtained in the first strategy coincides with the fidelity obtained in the second one, which shows that for the studied system, the optimization procedure can be efficiently performed via the second strategy that requires less computational time.

\section{Comparison, T = 0.5 \& 0.8}
\label{app_comparison}

\begin{figure}[ht]
    \centering     \includegraphics[width=\linewidth]{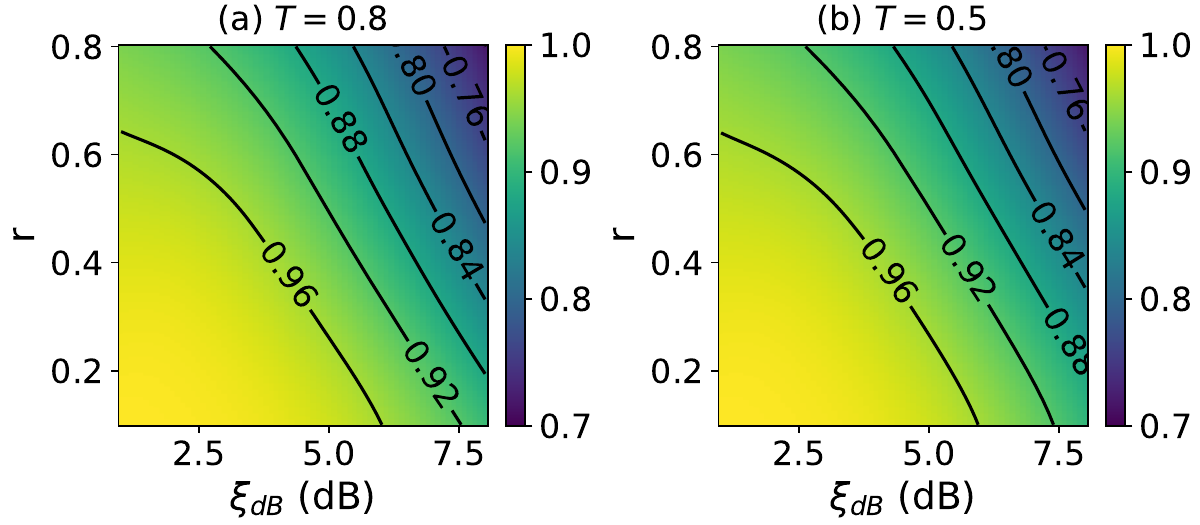}
    \caption{Optimized fidelity  for different combinations of $r$ and $\xi_{dB}$ values (target states) for the beamsplitter transmission coefficient  (a) T = 0.8 and (b) T = 0.5. }
    \label{fig:comparison}
\end{figure}

\newpage
\bibliography{refs}% Produces the bibliography via BibTeX.

\end{document}